\documentclass{mn2e}
\usepackage{epsfig}
\usepackage{times}
\begin{document}

\title[black hole binary jets] 
%{X-ray signatures of relativistic ejection events in black hole X-ray binary systems}
{Jets from black hole X-ray binaries: testing, refining and extending
empirical models for the coupling to X-rays} \author[Fender et al.]
{R. P. Fender$^{1,2}$\thanks{email:rpf@phys.soton.ac.uk}, J. Homan$^3$,
T.M.Belloni$^4$\\ $^1$ School of Physics and Astronomy, University of
Southampton, Highfield, Southampton, SO17 1BJ, UK\\ 
$^2$ Astronomical Institute `Anton Pannekoek', University of Amsterdam, Kruislaan 403, 1098 SJ Amsterdam, The Netherlands\\
$^3$ MIT Kavli Institute for Astrophysics and Space Research, Cambridge, MA, USA\\ 
$^4$ INAF-Osservatorio Astronomico di Brera, Merate (LC), Italy\\} 
\maketitle

\begin{abstract}
In this paper we study the relation of radio emission to X-ray
spectral and variability properties for a large sample of black hole
X-ray binary systems. This is done to test, refine and extend --
notably into the timing properties -- the previously published
`unified model' for the coupling of accretion and ejection in such
sources. In 14 outbursts from 11 different sources we find that in
every case the peak radio flux, on occasion directly resolved into
discrete relativistic ejections, is associated with the bright hard to
soft state transition near the peak of the outburst. We also note the
association of the radio flaring with periods of X-ray flaring during
this transition in most, but not all, of the systems. In the soft
state, radio emission is in nearly all cases either undetectable or
optically thin, consistent with the suppression of the core jet in
these states and `relic' radio emission from interactions of
previously ejected material and the ambient medium. However, these
data cannot rule out an intermittent, optically thin, jet in the soft
state.  In attempting to associate X-ray timing properties with the
ejection events we find a close, but not exact, correspondence between
phases of very low integrated X-ray variability and such ejections. In
fact the data suggest that there is not a perfect one-to-one
correspondence between the radio, X-ray spectral or X-ray timing
properties, suggesting that they may be linked simply as symptoms of
the underlying state change and not causally to one another.  We
further study the sparse data on the reactivation of the jet during
the transition back to the hard state in decay phase of outbursts, and
find marginal evidence for this in one case only. In summary we find
no strong evidence against the originally proposed model, confirming
and extending some aspects of it with a much larger sample, but note
that several aspects remain poorly tested.
\end{abstract}

\section{Introduction}

%\linenumbers 

Accreting stellar-mass black holes in binary systems display different
accretion `states', characterised primarily by different X-ray
spectral and variability properties (e.g. Nowak 1995; Homan et
al. 2001; Homan \& Belloni 2005, hereafter HB05; Remillard \&
McClintock 2006; Klein-Wolt \& van der Klis 2007; Done, Gierlinski \&
Kubota 2007; Belloni 2009). Understanding empirically the relation of
these states to the formation of powerful relativistic outflows, or
jets, is a key goal for more detailed theoretical understanding of the
coupled accretion:jet formation processes (e.g. Meier 2001; Livio,
Pringle \& King 2003; Ferreira et al. 2006). It also allows us to
estimate the kinetic feedback to the ambient medium (e.g. Heinz \&
Sunyaev 2002; Fender, Maccarone \& van Kesteren 2005), and to make
direct comparisons with supermassive black holes in active galactic
nuclei (e.g. Falcke \& Biermann 2001; Merloni, Heinz \& di Matteo
2003; Falcke, K\"ording and Markoff 2004; K\"ording, Jester \& Fender
2007; Fender 2008).

In Fender, Belloni \& Gallo (2004, hereafter FBG04; see also Fender \&
Belloni 2004) we outlined a unified picture for the disc:jet coupling in
black hole X-ray binaries, where 'disc:jet' should be taken as
shorthand for the relation between the infall and outflow of matter
around the compact object, in all its various proposed geometries (and
`disc' not taken to simply mean the optically thick, geometrically
thin, variety of accretion flow). This was based primarily on the
relation between X-ray and radio emission in four well-studied
systems. In very brief summary, although we recommend the interested
reader to read FBG04, the model's major components are:

\begin{itemize}
\item{A steady, powerful, relatively low bulk velocity (bulk Lorentz
  factor $\Gamma \leq 2$) jet is always present in the canonical hard
  X-ray state.}
\item{As a source makes a hard $\rightarrow$ soft X-ray state
  transition, the jet becomes initially unstable and then produces a
  major flare which in some cases has been directly resolved into a
  major relativistic ejection event (with $\Gamma \geq 2$). The point
  at which this occurs in the X-ray hardness-intensity diagram (see
  below) is referred to as the `jet line'.  The inferred origin for
  the flaring is internal shocking as this faster, transient jet runs
  into the pre-existing slower jet from the hard state.}
\item{Subsequently in the soft X-ray state the core jet is off, or at
  least much weaker.}
\item{In the soft $\rightarrow$ hard transition in the latter phases
  of the outburst the steady hard state jet reforms without a major
  flare (no slower jet in front to run into).}
\item{In trying to understand these observational conclusions, it was
  suggested that the `jet line' corresponded to the point at which an
  inwards-moving inner disc edge reached the innermost stable
  circular orbit (ISCO).}
\end{itemize}

Empirical aspects of this model were confirmed for a different system
in Corbel et al. (2004).  In this paper we take a far larger sample of
black hole X-ray binary outbursts and use them to test and refine the
model of FBG04 in the context of X-ray spectral states. In addition we
extend the model by making for the first time a comprehensive attempt
to include the X-ray short-timescale variability properties of the
systems.

\section{Data analysis}

\subsection{X-rays}

The X-ray data presented in this paper were obtained with the
Proportional Counter Array (PCA) onboard the {\it Rossi X-ray Timing
Explorer} (RXTE). In our analysis we make extensive use of the
hardness-intensity diagram (HID) as an indicator of the X-ray spectral
state of a black hole binary in outburst. The HIDs presented here were
taken from Homan et al.\ (in prep.). They were constructed from {\tt
standard2} mode data from the PCA. These data were corrected for
background, but not for dead time (typically a few percent).  Averaged
count rates were extracted for each observation in three bands:
channel 20-40, 3-10, and 1-129, roughly corresponding to 9.5--17.0
keV, 2.8--5.5 keV and 2--60 keV, respectively. The ratio of the count
rates in first two bands was used as an indicator of the spectral
hardness and the third band as the broadband intensity.  In case fast
intensity or hardness variations were observed within an observation,
the observation was split into two or more parts and hardness and
intensity were calculated for each individual part.

In addition to HIDs we also studied the X-ray variability properties
of all sources. Again, these data were taken from Homan et al.\ (in
prep.). Power spectra were constructed from high-time-reolution data
from the Proportional Counter Array (PCA), using most of
the PCA energy range and following standard fast-fourier techniques
(see e.g.\ Homan et al. 2005 for a detailed
description). As a measure of the strength of the X-ray variability we
extracted the rms-normalized power from the 0.01--64 Hz frequency range,
for each observation (or observation segment).

\subsection{Radio}

Radio data presented in this paper were gathered exclusively from the
literature, and references are provided in the brief descriptions for
each outburst in section 2.

\section{Radio flares and ejection events in the context of outburst evolution}

As noted above, in the following analysis we use Hardness-Intensity
Diagram (HID) as an indicator of the X-ray spectral state of a black
hole binary in outburst. This is the framework within which the
analyses of FBG04, HB05 and others were also set. In the HID the
vertical axis simply corresponds to X-ray count rate within some band,
and the horizontal axis to the `hardness', or ratio of counts measured
in two narrower sub-bands, such that `harder' spectra (those with a
relatively higher proportion of high-energy to low-energy X-rays
within the band) are on the right hand side, and `softer' spectra are
on the left hand side. Black hole and neutron star binaries (and maybe
also white dwarfs) are known to demonstrate hysteresis in the tracks
they trace in such diagrams during outburst, typically making hard
$\rightarrow$ soft transitions at higher luminosities than the return
soft $\rightarrow$ hard transition (FBG04; HB05; Koerding et
al. 2006). Section 2.1 (above) gives the X-ray bands used for the
hardness ratios in this paper.

In Fig 1 we present HIDs for 14 outbursts from 11 different black hole
X-ray binaries, annotated with symbols indicating times of radio
detections, radio flares, and times when the radio counterpart was
undetectable. Note that we do not discriminate between levels of radio
emission in these figures: it is either detected, a special case of
which is the peak flux, or it is not. Nevertheless this is enough to
give us a good test of the scenario presented in FBG04. In the
following we describe the systems and their outbursts individually, in
the order in which they are presented in Fig 1. Note that while there
are more RXTE data on black hole outbursts than those which are
presented here, (i) these were all the outbursts for which we had
useful X-ray and/or radio coverage, (ii) we are not aware of any other
data set of results which contradict the empirical conclusions drawn
here.

\subsection{GRO J1655-40}

This black hole binary is famous as the second known superluminal
source in the galaxy (Tingay et al. 1995; Hjellming \& Rupen
1995). The 1994 outburst with which these superluminal ejections were
associated was not covered by RXTE (which was launched in December
1995), but two subsequent outbursts, in 1996 and 2005, have been well
covered in X-rays.

\subsubsection{1996}

Radio coverage of this outburst was poor, but comparing a VLA
non-detection ($<3$mJy) and a subsequent MOST detection at level of
$55 \pm 5$ mJy (843 MHz) indicates that a radio flare (i.e. probable
ejection event) occurred somewhere between 1996 May 20-28 (MJD
50223-50231). This radio flare occurred about half way in the
transition between hard and soft X-ray states.

\subsubsection{2005}

The radio and RXTE coverage of this outburst was much better than in
1996. The source traces a clear pattern in the HID. Radio
observations with the VLA were reported in Rupen, Dhawan \&
Mioduszewski (2005a,b,c,d,e,f,g,h). The relatively weak radio peak
occurred during a phase of X-ray flaring during the hard
$\rightarrow$ soft transition.

\subsection{4U 1630-47}

The recurrent black hole transient 4U 1630-47 has undergone five
outbursts between 1996 and 2005.  The source has been identified with a
weak radio counterpart (Hjellming et al. 1999) but is not always
detected (Gallo et al. 2006). In Fig 1 we present a HID for the 1998
outburst of the source, which has the best radio coverage (Hjellming
et al. 1999). As with most sources, a jet has not been directly
resolved in this source, so we can only rely upon the peak flux as an
approximate indicator of an ejection event. This peak flux occurs
about half way, in terms of relative X-ray spectral hardness, between
the canonical hard and soft (or thermal dominant) X-ray states, around
MJD 50861. By back-extrapolating the radio light curve, Hjellming et
al. (1999) estimated a probably ejection date of $\sim$MJD 50851, ten
days earlier than this flare.

\begin{figure*}
\centerline{\epsfig{file=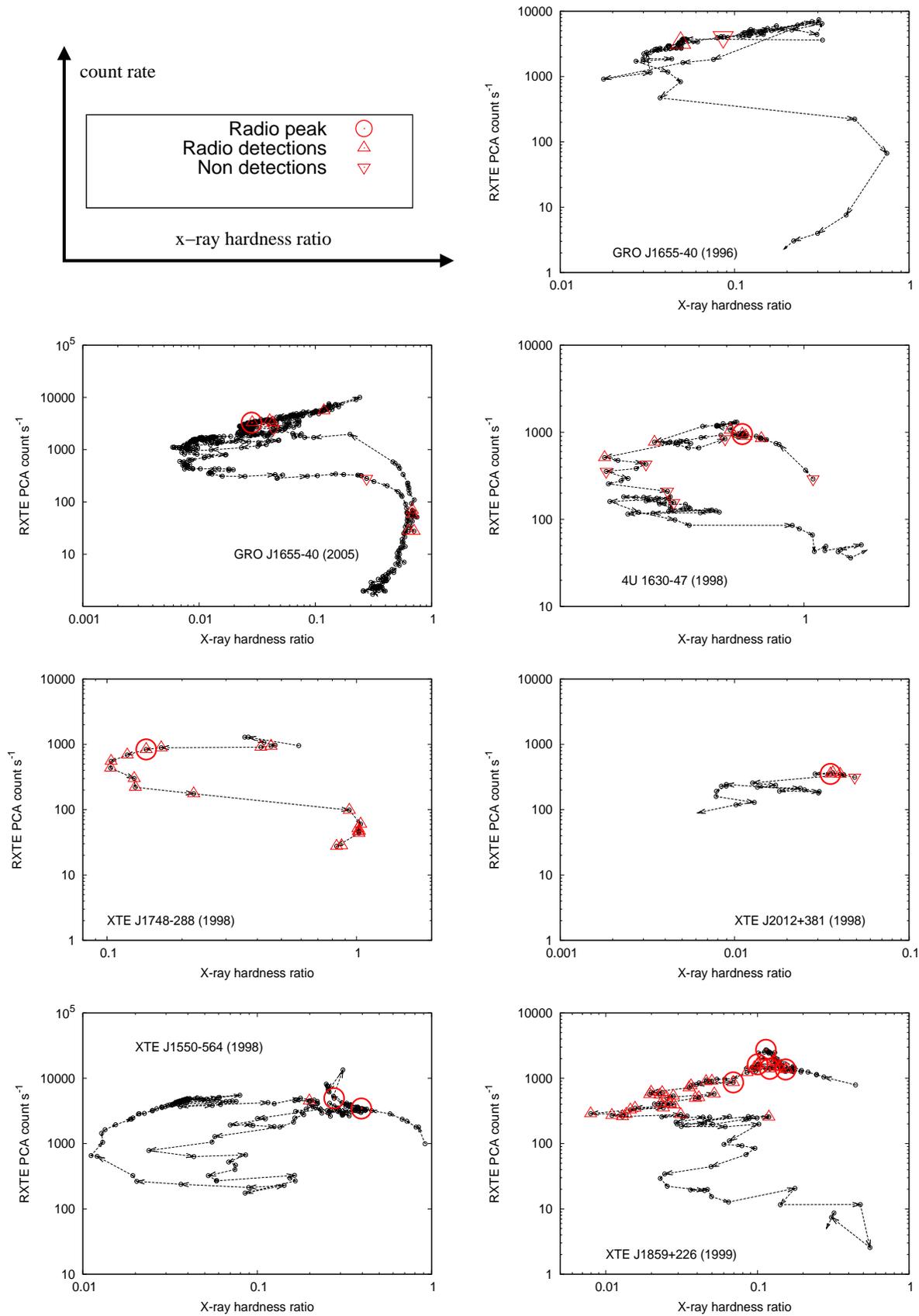, width=16cm, angle=0}}
\caption{(a) Hardness intensity diagrams for black hole X-ray binary
  outbursts. The key at the top left indicates the symbols for radio
  detection, radio peak, and radio upper limit, which are plotted on
  top of the HIDs at the location of the nearest RXTE X-ray
  observation, usually within 24 hr. In some cases there are clear
  distinct radio peaks which probably correspond to individual
  ejection events, and so multiple radio peaks are indicated.}
\end{figure*}

\begin{figure*}
\setcounter{figure}{0}
\centerline{\epsfig{file=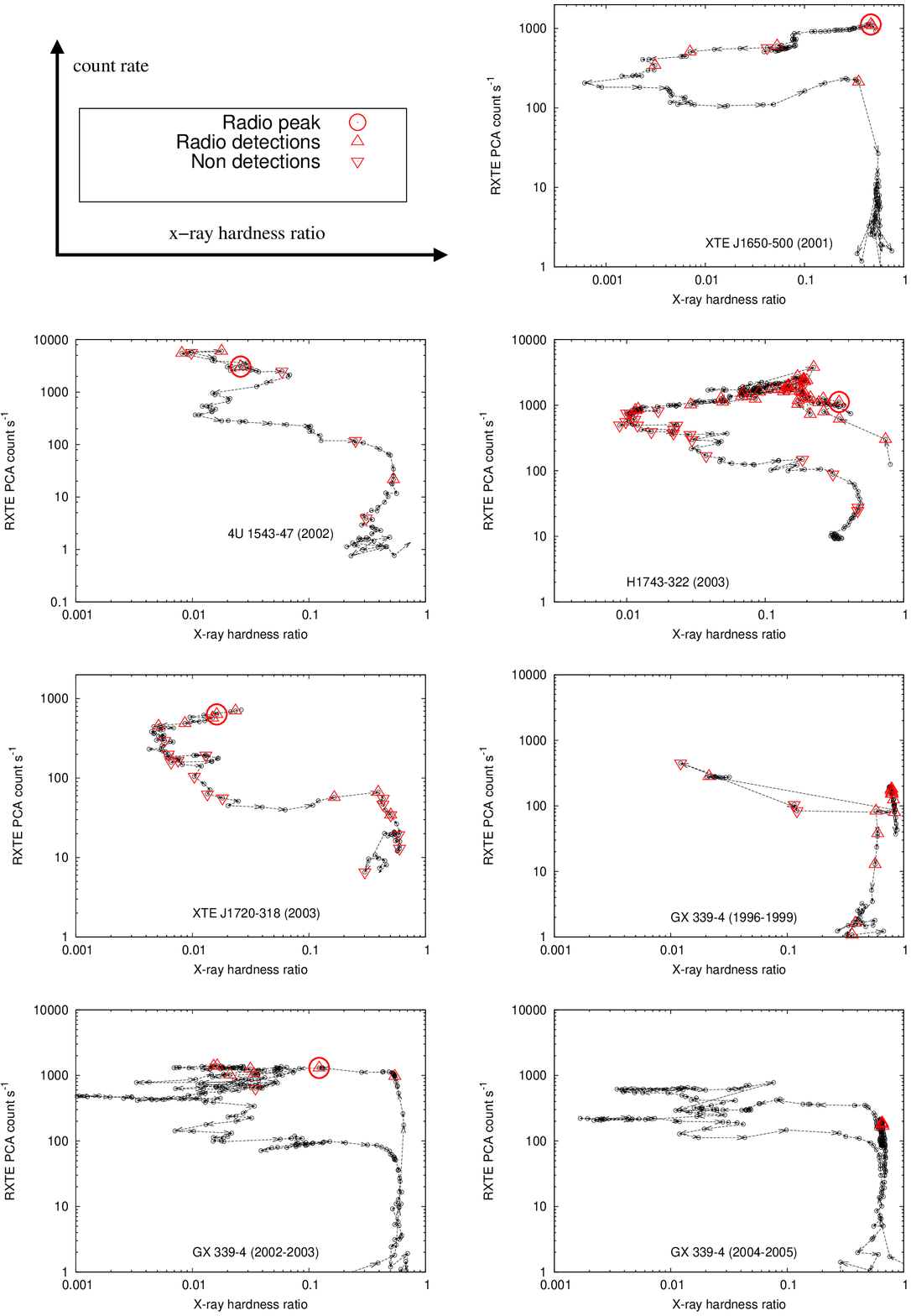, width=16cm, angle=0}}
\caption{(continued)}
\end{figure*}

\subsection{XTE J1748-288}

This source underwent a bright and prolonged X-ray outburst with
complex radio behaviour in 1998. Analysis of the radio flux and images
has recently been undertaken in Brocksopp et al. (2007) and
Miller-Jones et al. (prep). The peak radio flux is clearly during the
hard $\rightarrow$ soft transition, but radio emission was essentially
detected every time the source was observed throughout the entire
period of the bright X-ray outburst. Observations with the VLA
(Hjellming, Rupen \& Mioduszewski 1998; Hjellming et al. 1998b;
Miller-Jones et al. in prep) clearly show that a lot of this is due to
resolved component which appear to interact with the environment
and/or each other. Note that extrapolating the early proper motions of
the resolved maps in Miller-Jones et al. (in prep) indicates that the
first ejection every took place before the first X-ray observations,
which may in principle have still been in the canonical low/hard
state. The equally plausible alternative is that the initial proper
motions were much higher, and the ejecta decelerated as they
interacted with the ISM (as in e.g. XTE J1550-564 -- Corbel et
al. 2001).

\subsection{XTE J2012+381}

XTE J2012+381 is a black hole candidate X-ray transient that went
into outburst in 1998 (e.g. Campana et al. 2002). The RXTE coverage is
sparse. Hjellming, Rupen \& Mioduszewski (1998c) and Pooley (1998)
report on a weak but variable radio counterpart. The variability
reported by Pooley (1998) may indicate an ejection around May 30 (MJD
50963)

\subsection{XTE J1550-564}

XTE J1550-564 is a fascinating source with a wealth of interesting
properties in terms of its disc--jet coupling, including powerful,
moving, X-ray jets (Corbel et al. 2002) and large loops in the HID
(Fig 1). Hannikainen et al. (2001) report VLBI images of relativistic
jets following a major flaring event around MJD 51078, with strong
evidence for a secondary core ejection event around MJD 51080; Homan
et al. (2001) report a bright, optically thin radio source around MJD
51248.  These dates are indicated in Fig 1. Note that there several
unpublished radio observations of this outburst which may in the
future shed some further light on the evolution of this outburst.

\subsection{XTE J1859+226}

XTE J1859+226 is a source for which there was both good X-ray and
radio coverage around the brightest phases of the outburst (see
e.g. Casella et al. 2004 for an X-ray study of the outburst).  Radio
data for this bright outburst are from Brocksopp et al. (2002), who
noted five radio flare events in a sequence of declining strength
(reminiscent of those seen in GRS 1915+105 e.g. Fender et al. 1999).
All five radio flare events occur during a prolonged period of X-ray
flaring around the middle of the hard to soft transition, with the
strongest of these happening at the time of a local peak in the X-ray
intensity. Fig 2 presents a close-up of the HID around the time of
these five flare events. All of these events occur around the same
hardness, with the greatest outlier being revealed as associated with
a brief and rapid hardening of the X-ray spectrum of total duration
$\leq 1.8$ days. This is explored in more detail in the discussion
(see Fig 2).

\subsection{XTE J1650-500}

Corbel et al. (2004) present a detailed discussion of the ATCA radio
observations of this source during its 2001 outburst. The peak radio
flux is detected close to the start of the hard $\rightarrow$ soft
transition, although there is subsequently a lack of radio
observations for 15 days and the true peak may have been
missed. Subsequently the radio emission is found to be temporarily
undetectable, but then to re-emerge in the soft state.

%\subsection{XTE J1908+094}
%This faint X-ray transient had good coverage with VLA both during the
%outburst and at fainter level (Jonker et al. 2004).
%** Rupen's webpage reports resolved jet and a number of detections and
%upper limits. Need to seek permission to extract the results used in
%the plot.  The peak radio flux occurred near the start of the hard to
%soft transition, followed by some non-detections and subsequent weak
%re-detection. **
% HAVE HAD NO LUCK GETTING PERMISSION SO WILL NOT USE

\subsection{4U 1543-47}

Park et al. (2004) and Kalemci et al. (2005) report in detail
multiwavelength observations of this recurrent transient. In
particular, Park et al. (2004) detected a rapidly rising radio flare
on MJD 52443. During the decay of the outburst, radio emission is
observed to reactivate by MJD 52487, a week after the MJD 52480 date
that Kalemci et al. (2005) report for the soft to hard state
transition, based on timing as well as spectral properties.

\subsection{H 1743-332}

The 2003 outburst of this recurrent transient received excellent radio
and X-ray coverage, as reported in McClintock et al. (2006). The HID
reveals a peak in the radio emission near the start (i.e. hard side)
of the hard to soft transition (as in XTE J1859+226 this is associated
with a phase of X-ray flaring), with many more detections in the
intermediate and initial soft states before a large number of upper
limits in the later soft state and during the soft to hard transition.

\subsection{XTE J1720-318}

Brocksopp et al. (2005) report radio coverage of this outburst. The
peak observed radio emission occurs around MJD 52655, at which point
the source is already in a rather soft state.  Subsequently there are
a number of non-detections in the soft state.  Brocksopp et al. (2005)
also report that the radio emission in this source appears to switch
back on between MJD 52715--52728 (increasing by a factor of at least
two despite a more or less steady level of X-ray emission), with two
further detections at comparable levels. This may be the best case to
date for the reactivation of the jet during the soft to hard
transition (see further discussion below).

\subsection{GX 339-4}

GX 339-4 has been a key source in our understanding of the relation
between X-ray luminosity / states and radio emission (e.g. Hannikainen
et al. 1998; Fender et al. 1999; Corbel et al. 2000, 2003; Gallo et
al. 2004). This is due in part to its regular bright excursions and
very clear hysteretical tracks in the HID (e.g. Homan \& Belloni 2005;
Belloni et al. 2003; Belloni 2009).

\subsubsection{1996--99}

This was the outburst which clearly established the dramatic decrease
in the X-ray:radio ratio during soft X-ray states. Radio observations
are reported in Fender et al. (1999) and, more comprehensively, in
Corbel et al. (2000).  No clear bright radio flare was observed,
although Fender et al. (1999) report a brief period of optically thin
radio emission arising between MJD 50828 and 50840. The comparison of
RXTE ASM and BATSE X-ray data with the radio monitoring clearly
indicate the suppression of the radio emission in the soft state,
although the poor RXTE PCA coverage means that this is not so clear in the
figure presented here.

\subsubsection{2002--03}

The 2002--03 outburst provided a clear and precise detection of a
radio flare and subsequent large-scale jet (Gallo et al. 2004). That
this flare occurred during the hard to soft transition was already
established in FBG04, and is shown again in Fig 1. Once again the
flare occurred about half way through the transition from hard to soft
X-ray states. Subsequent radio brightenings were thought to be
associated with the formation of shocks in the large-scale jet,
something which may account for radio emission in the soft state in
several of the other systems.

\subsubsection{2004--05}

The 2004--05 outburst of GX 339-4 did not receive much radio coverage,
with Miller et al. (2006) publishing the only radio
detection. However, the X-ray coverage was excellent, so it is worth
presenting.

\section{The moment of major ejection in the hardness intensity diagram}

\begin{figure}
\centerline{\epsfig{file=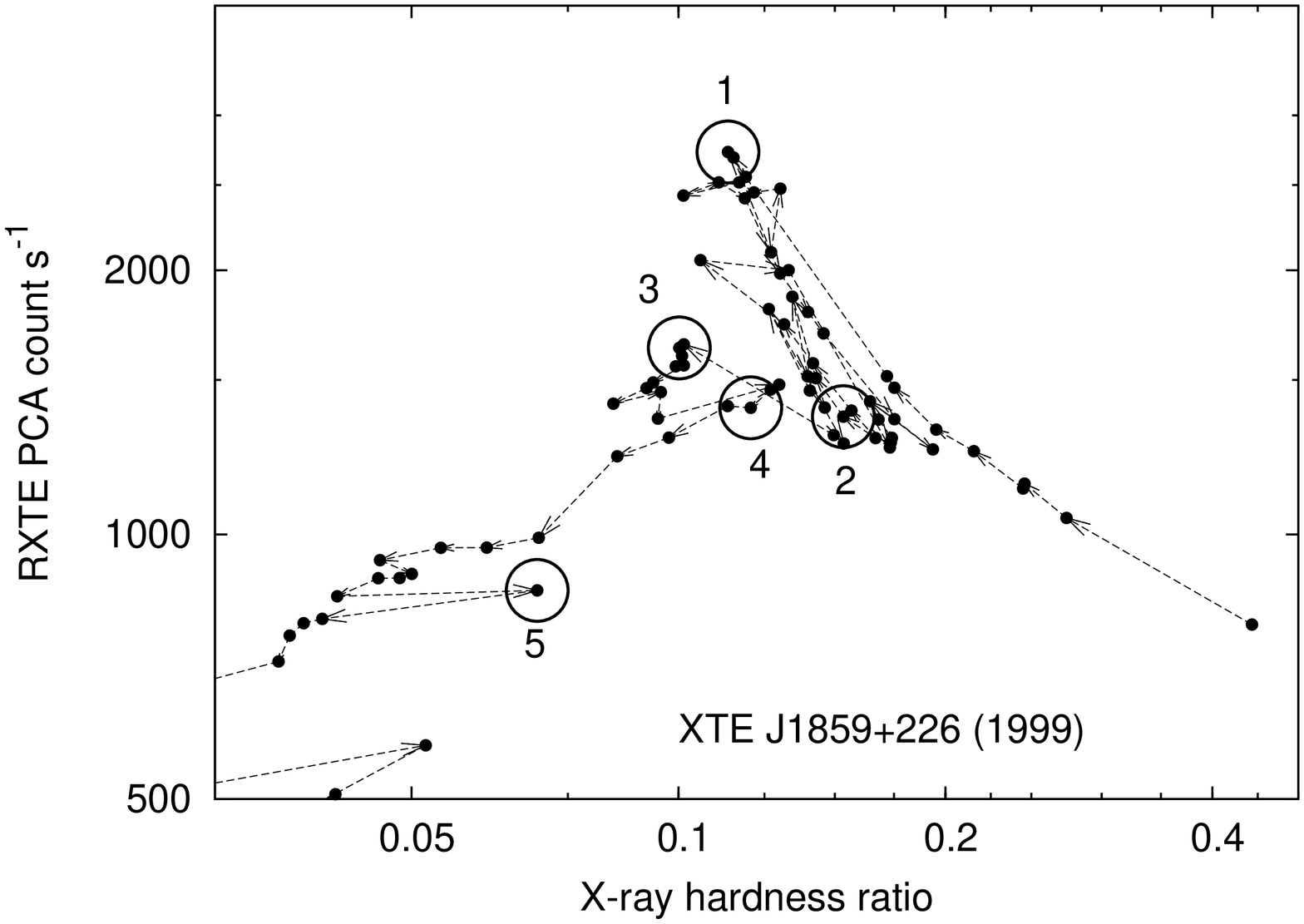, width=7cm, angle=270}}
\caption{Closer inspection of the period of repeated flaring in XTE
  J1859+226 in 1999. Only the five flare peaks from Brocksopp et
  al. (2002) have been indicated, not all the radio detections (see
  Fig 1). Arrows indicate the direction of motion in the HID; numbers
  indicate the sequence of events and correspond to those in Fig 4 of
  Brocksopp et al. (2002).  All of the flare events are associated
  with a very similar X-ray colour. The apparent outlier at the
  softest colour (flare 5), is revealed to be associated with a very
  rapid (total duration $\leq$ 1.8 days) excursion to a harder state.}
\label{1859zoom}
\end{figure}

In the scenario of FBG04, the major ejection event in a black hole
transient outburst occurs during the hard to soft state transition. In
all cases presented in Fig 1 the radio flare and/or resolved ejection
event occurs between the brightest hard state and the transition to
the soft state, in the 'intermediate states' of Homan \& Belloni
(2005). This confirms this aspect of the FBG04 model, with a much
broader sample.

However, it is important to note that several systems, e.g. XTE
J1550-564, XTEJ1859+226, H1743-322, show extended periods of strong
X-ray flaring during the hard $\rightarrow$ soft transition, and that
it is {\em these} phases which seem to be associated with the
brightest radio flares. Our archetypal varying black hole, GX 339-4,
it seems, is actually a notable exception to this pattern, in showing
a relatively smooth hard $\rightarrow$ soft transition without phases
of strong X-ray flares.

Nevertheless, we can ask some naive questions and see where they lead;
for example does the radio flaring always occur at the same hardness ?
It is hard to compare the evolution of different sources with
different hardness ranges. In comparing different sources with similar
absorption (GX 339-4 and XTE J1650-500) we find no evidence for
ejections always occuring at the same hardness. Timing studies (see
below) or comparison to the Disc Fraction Luminosity Diagram
(Koerding, Jester \& Fender 2007; Dunn et al. 2008 and in prep) may
provide clearer comparisons between different systems. In addition, to
further refine the connection in the future we will need high angular
resolution and/or high frequency radio/mm/IR observations in order to
avoid uncertainty associated with initially large optical depths.

We may also study in detail the hardness evolution of a single source
when there is evidence for multiple ejection events. Fig 2 highlights
the region of the HID for XTE J1859+226 in which it seems that five
discrete ejection events occurred (Brocksopp et al. 2002). The first
four events occur in a narrow range of hardness $0.1 \leq H \leq
0.2$. The fifth event, which in Fig 1 looks like something of an
outlier, is revealed in Fig {\ref{1859zoom}} to be associated with a
rapid (total duration $\leq 1.8$ days) excursion back to a harder
X-ray state. So, these data support the suggestion that the all five
flare events occur when the source is in a very similar spectral
state. Are all the events associated with hard to soft transitions
(i.e. left to right motion in the HID) ? This is harder to
say. Certainly most seem to be associated with some change in
direction or (2D) excursion, but the motion during this period is so fast
that it is very hard to tell. Note that Casella et al. (2004) find
that this region of the HID is very rich in strong quasi-periodic
oscillations (QPOs), and provide more details of the QPO
phenomenology.

The study of GX 339-4 reported in Belloni et al. (2005) implies that
the lower-luminosity outburst of GX 339-4 in 2004--5 made the 'hard
intermediate state' (HIMS) to 'soft intermediate state' (SIMS)
transition at a larger hardness ratio than the outburst in
2002--03. This conclusion was reached based on X-ray spectral and
timing properties. If the moment of major ejection is associated with
this transition, then the `jet line' may be diagonal, or more complex,
in the HID (see discussion later). It is also worth reminding the
reader that while we do not discuss this source specifically here, the
large number of events observed from the powerful jet source GRS
1915+105 (Fender \& Belloni 2004) are all, to our knowledge,
consistent with the patterns described in FBG04 (see also Soleri,
Belloni \& Casella 2008 for the association of type-B QPOs with
ejection events in this source).

\section{X-ray variability as an indicator of ejection events}

A complementary view of X-ray binary states and state transitions is
provided by studying the X-ray variability of such sources, typically
on timescales of mHz to kHz, via Fourier power spectra (for recent
reviews see van der Klis 2006; Remillard \& McClintock 2006;
Klein-Wolt \& van der Klis 2007; Belloni 2009). These X-ray timing
properties were not considered, however, in anything but the simplest
sense in the model of FBG04. It has since been noted by several
authors that the sharpest transition in the combined spectral and
temporal classification of states is a sharp transition in timing
properties between the 'Soft Intermediate State' (SIMS) and the 'Hard
Intermediate State' (HIMS) (e.g. Belloni et al.  2005; Casella,
Belloni \& Stella 2005; Klein-Wolt \& van der Klis 2007).  Following
the model presented in FBG04, several authors have speculated that
there may be a close link between the sharp X-ray timing state
transition and major relativistic ejection events (e.g.  Ferreira et
al. 2006; Klein-Wolt \& van der Klis 2007). In the following we will
explore the possibility that properties of the rapid X-ray variability
may be a better indicator of ejection events than the energy spectrum.

In many cases analysis of rapid variability is a question of detailed
fitting to X-ray power spectra, and of often subjective
interpretation.  Homan et al. (in prep) have identified `zones' of
significantly and rapidly reduced X-ray r.m.s. variability in the
noise continuum of the power spectrum, which occur in the intermediate
X-ray states (see Belloni et al. 2004 for a clear example in the case
of GX 339-4). In fact, the transition from the HIMS to the SIMS, the
most dramatic change in timing properties, is well indicated (in some
interpretations, {\em defined}) by a dramatic drop in the integrated
X-ray r.m.s. variability. This drop in r.m.s. variability is often,
but not exclusively (as far as the data currently reveal) associated
with 'type B' QPOs as defined by Casella et al. (2005). Characteristic
X-ray power spectra throughout a hard $\rightarrow$ soft state change
in GX 339-4 are presented in Fig 3 (alphabetical order corresponds to
temporal order), and their correponding temporal and
hardness-r.m.s. diagrams (HRDs) presented in Figs 4 and 5 (power
spectra over the preceding 30 days, for GX 339-4, are presented in
Klein-Wolt and van der Klis 2008). Power spectra 'c', 'd' and 'e'
correspond to the low-r.m.s. zone; the dramatic transition from the
strongly variable power spectrum 'b' to the low-r.m.s. zone in 'c'
takes place in $\leq 26$ hr (in fact the transition has been observed
to be as rapid as $\leq 32$ sec -- e.g. Casella et al. 2004).  It is
worth noting that immediately following these zones and phases with
very strong QPOs (e.g. power spectrum 'd') the power spectrum returns
to the band limited noise shape which it had in the HIMS prior to the
zone, however with a signficantly softer spectrum. This would seem to
set some limit on the connection between the hard state noise and the
jet launching mechanism, given that the jet properties appear to
change dramatically following the major flare event (jet is suppressed
and/or becomes optically thin; see next section). Klein-Wolt \& van
der Klis (2008) make a similar point in that frequencies associated
with QPOs and other features in X-ray power spectra appear to evolve
quite smoothly in frequency across the intermediate state transitions,
at the same time that the jet appears to flare and then disappear.
Note that the low-r.m.s. zones are sometimes 'filled in' in the HRDs
(Fig 5) during the hard $\rightarrow$ soft return state transition.

\begin{figure}
\centerline{\epsfig{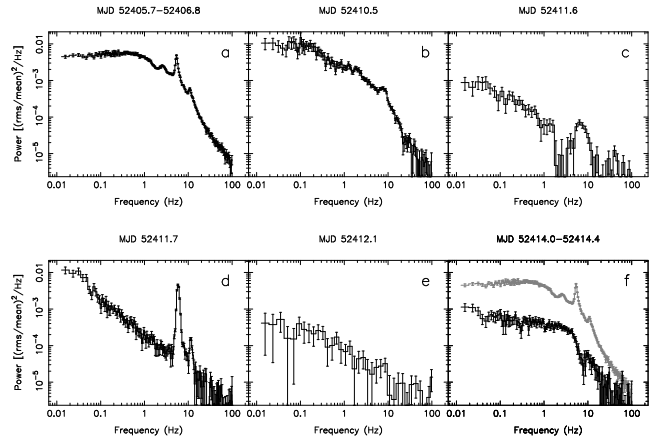}}
\caption{Detailed evolution of the X-ray power spectra of GX 339-4
before and after the radio flare and subsequent low-r.m.s. `zone' in
the 2002/3 outburst. The epochs corresponding to the letters are also
indicated in Figs \ref{3panels} and \ref{zones}. The power spectrum
from panel a is replotted for reference in panel f.}
\label{power}
\end{figure}

\begin{figure}
\centerline{\epsfig{file=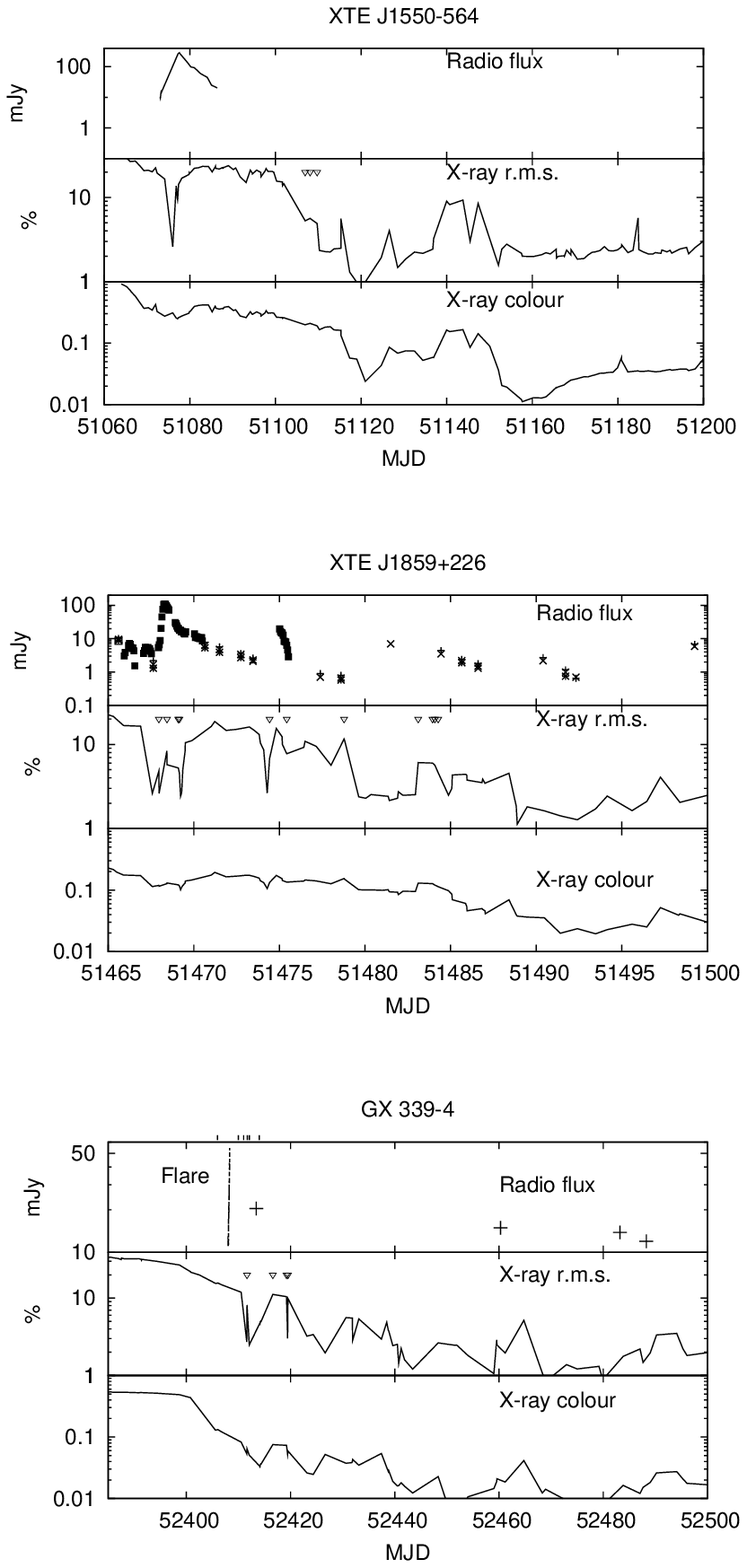, width=8cm, angle=0}}
\caption{Evolution of radio flux density, (log) r.m.s. variability and
(log) X-ray colour with time for the major outbursts of three black
hole transients. Inverted open triangles in the r.m.s. panels indicate
times when type-B QPOs were observed.  In XTE J1550-564 the major
radio flare occurs just after a very sudden drop in X-ray variability
(which itself corresponded to a very strong flare in total X-ray
flux). In GX 339-4 the one observed bright radio flare occurs days
before the clearest dip in the X-ray variability; the sequence of
small ticks at the top of the figure indicates the epochs of the six
power spectra presented in Fig \ref{power}.  In XTE J1859+226 a
sequence of radio flares appear roughly correlated with drops in
r.m.s. variability superposed on a declining trend which seems well
matched to the softening X-ray colour. An alternative view of these
events, in the hardness--r.m.s.  plane, is presented in Fig
\ref{zones}.}
\label{3panels}
\end{figure}

\begin{figure}
\centerline{\epsfig{file=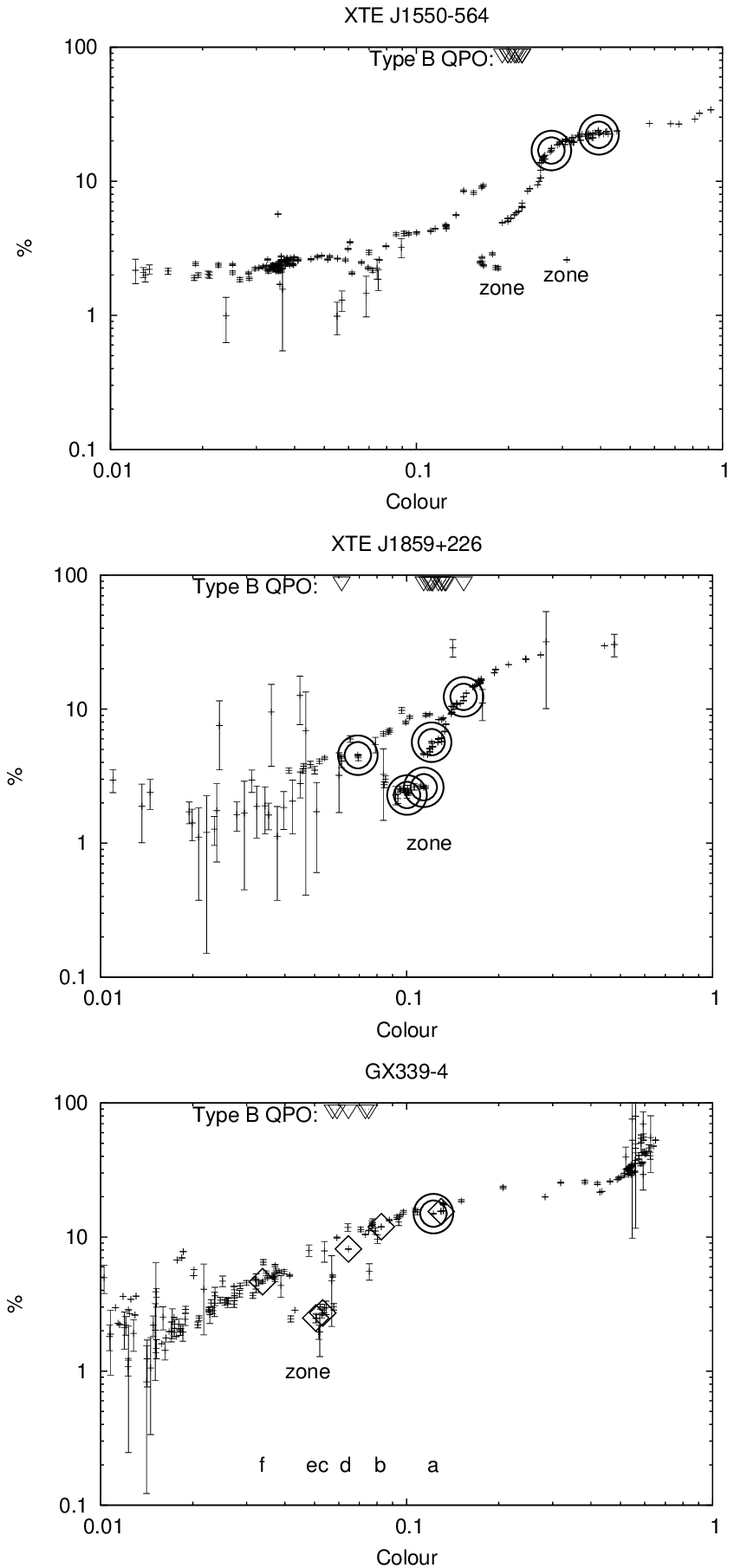, width=8cm, angle=0}}
\caption{Hardness-r.m.s. diagrams (HRDs) for the three transients
investigated in detail in Fig \ref{3panels}, with moments of radio
flares indicated by the concentric rings. In the case of GX 339-4, the
locations in the diagram of the six power spectra presented in Fig
\ref{power} are indicated by diamonds, and the sequence indicated by
the letters towards the lower edge of the figure. Inverted open
triangles indicate the presence of type-B QPOs.}
\label{zones}
\end{figure}

In the following section we study in detail the radio flux density,
X-ray hardness (colour) and X-ray r.m.s. variability in detail for
three outbursts: XTE J1550-564 (1998), XTE J1859+226 (1999) and GX
339-4 (2002). This is our first detailed and quantitative attempt to
connect the sharp timing transitions with the exact moment of major
relativistic ejection. In Fig \ref{3panels} we present radio flux
density, X-ray r.m.s. and X-ray hardness against time for these three
outbursts, and in Fig \ref{zones} we plot the points of major radio
ejection (radio flares; exactly equivalent to the points of peak radio
flux in Fig 1) in the r.m.s. vs. colour diagram (see also Belloni et
al. 2005 for the r.m.s. vs. colour diagram for GX 339-4). In addition,
in both sets of figures we indicate, with inverted open triangles, the
times at which type-B QPOs have been observed.  Taken together, Figs
\ref{3panels} and \ref{zones} indicate a connection between X-ray
spectral and timing changes and points of major ejection which looks
promising at first but may weaken under closer inspection.  We note
that we only have coverage of the full transition from canonical hard
to soft states in a subset of the outbursts presented in Fig 1; in
most cases this is due to the X-ray coverage having missed the initial
hard state. Russell \& Fender (2007) note that the timescale from
suppression of the optical/infrared flux (probably indicating the
beginning of the state transition) to radio flaring is typically 7--12
days.

\subsection{XTE J1550-564}

In the case of XTE J1550-564, the major radio flare and relativistic
jets reported in Hannikainen et al. (2001) seem to be clearly
associated with a very strong dip in the integrated X-ray r.m.s.,
which corresponded to a major $\sim 6$ Crab X-ray flare. The X-ray
peak occurs on MJD 51075--6, and the radio peak on 51078, consistent
with the optical depth-evolution of a powerful ejection (e.g. van der
Laan 1966; see also Discussion). The X-ray power spectrum at this peak
showed a very strong drop in r.m.s., with a prominent 'type C'
(Casella, Belloni \& Stella 2005; Sobczak et al. 2000) and possibly a
weaker 'type B' QPO (Belloni et al. in prep; not indicated in Figs 4
and 5 as a confirmed detection).  Spectral fits, combined with a clear
peak in the frequency of the low-frequency QPO at $\sim 13$ Hz,
suggest that for a transient phase the disc {\em may} reached a local
minimum in radius, although there are certainly alternative
explanations (see e.g. Sobczak et al. 2000). Note that type-C QPOs are
sometimes observed to reach higher frequencies (in both this and other
sources) and so even if the disc did reach a local minimum in radius,
there is no strong case that this was at the ISCO (this is of some
importance compared to the suggestion in FBG04 that major ejections
take place when the disc reaches the ISCO).  Regardless of the precise
details, the disc component was clearly very luminous and very hot at
this point (comparable, briefly, to GRS 1915+105), and the event is
strongly implicated in the production of a very powerful jet which was
still observable accelerating electrons to TeV energies in the ISM
some four years later (Corbel et al. 2002).  However it is important
to note that the drop in r.m.s. variability during this, and other,
zones cannot be ascribed solely to dilution of the signal by a
non-variable disc component.  Taking this bright event as an example,
both the disc and power-law components increased in luminosity by a
factor $\sim 3$, whereas the r.m.s. variability dropped by a factor of
$\sim 10$.  This event is clearly significantly harder than the later,
broader, zone of reduced r.m.s. (Fig 5).

\subsection{XTE J1859+226}

In XTE J1859+226 the sequence of five radio flare events appears to be
well correlated with local minima in both X-ray timing and colour. The
HRD clearly shows that all five reported events occurred around the
reduced r.m.s. zones. Casella et al. (2004) report the evolution of
the timing properties throughout this outburst in some detail.

\subsection{GX 339-4}

In GX 339-4 the strong radio flare reported by Gallo et al. 2004
clearly occurs around the end of the rapid X-ray colour change, but
clearly {\em before} the zone of reduced X-ray r.m.s. This is very
important measurement: unless we have missed a major radio event
(which is possible given the poor radio coverage in most cases) then
the radio ejection in GX 339-4 began at least two days before the
strong X-ray timing state transition, and also before the appearance
of 'type B' QPOs (Casella et al. 2005) which have also been suggested
to be associated with ejections (e.g. Soleri et al. 2008). The timing
behaviour during this outburst is summarised in Belloni et al. (2005).

\section{Radio emission in the soft state}

In a large number of the outbursts presented in Fig 1, there is
clearly some radio emission, at least initially, in the soft X-ray
states which generally follow the radio flare / ejection event. It is
unfortunate that the large number of upper limits to the radio
emission during the 1996--1999 soft state of GX 339-4 (Fender et
al. 1999; Corbel et al. 2000) were not accompanied by intensive RXTE
PCA monitoring, and so are not represented here.  Nevertheless, some
systems clearly show a lack of radio emission in the soft state once
the large flare has faded (see e.g. 4U 1630-47, H1743-322, XTE
J1720-318 in Fig 1).

\begin{figure}
\centerline{\epsfig{file=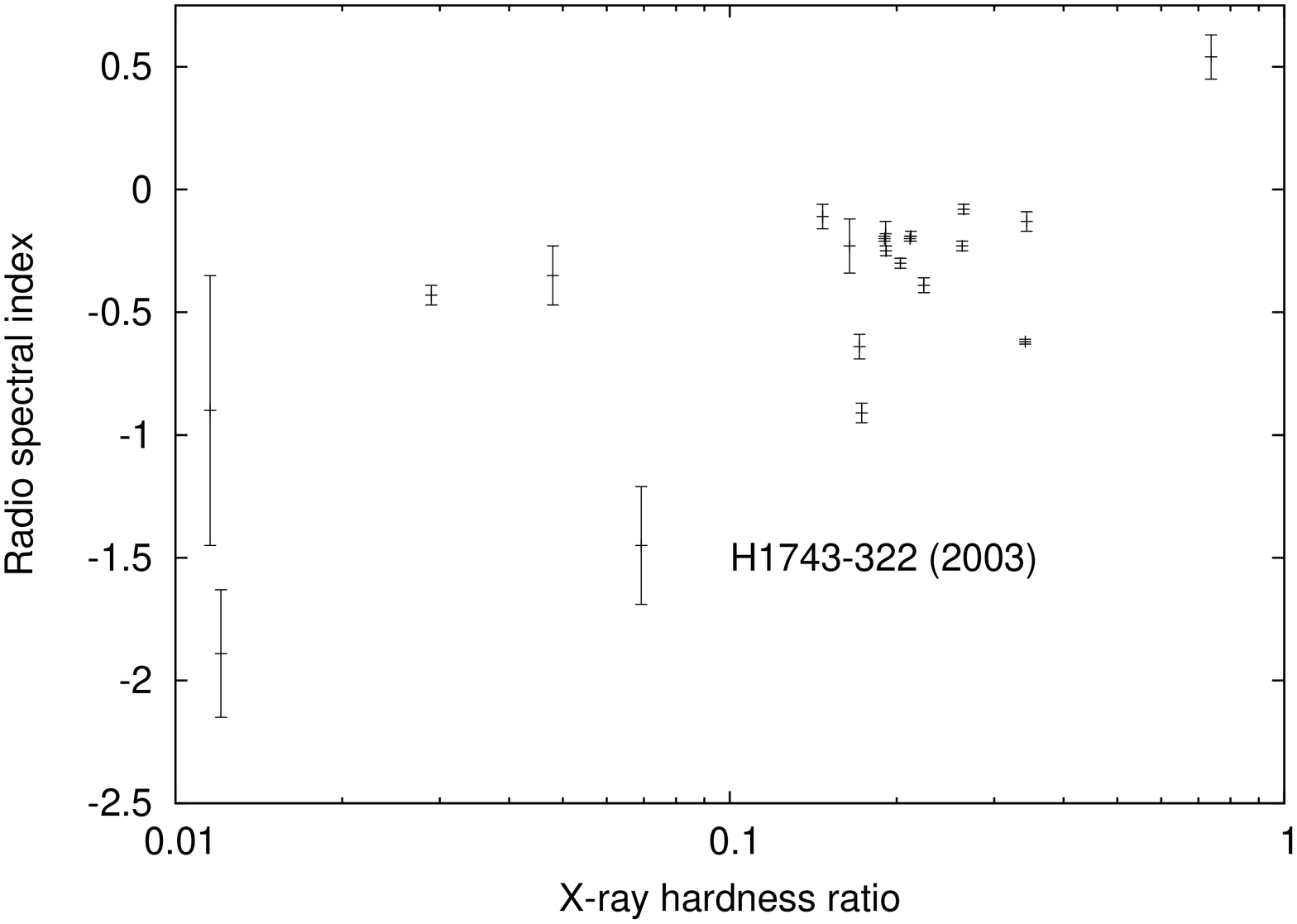, width=6cm, angle=-90}}
\caption{Radio spectral index as a function of X-ray colour for
H1743-322, based on the observations reported in McClintock et
al. (2006). There is a clear trend for more negative spectral indices,
indicating optically thin emission, in softer states. This is
consistent with, but not proof of, the suggestion that in the soft
state radio emission is associated with spatially extended jets and
shocks, and not with the core.}
\label{spindex}
\end{figure}

What are we to make of the radio detections in the soft state, given
that the current paradigm is that the jet is suppressed in this state
? There seem to be two alternatives:

\begin{itemize}
\item{{\bf I.} The core jet {\em is} strongly suppressed, and the radio
  emission is associated with rebrightenings at shocks as the jet propagates away from the binary.}
\item{{\bf II.} The core jet, in some cases, remains on, at least for a while, in the soft state}
\end{itemize}

We note that scenario I -- radio brightening which images reveal to be
displaced from the binary core -- has been directly observed in
several cases (e.g. Corbel et al. 2002; Gallo et al. 2004). In this
scenario, the radio emission we observe in the soft state should in
most cases have an optically thin radio spectrum, in contrast to the
flat radio spectrum associated with the hard state (Fender 2001). In
Fig \ref{spindex} we plot the radio spectral index for observations of
H1743-322 as a function of X-ray hardness, which support a trend
towards more negative spectral indices, indicative of a lower optical
depth, with decreasing hardness. In addition such emission should be
observed to monotonically fade as the ejecta expand in the surrounding
medium. Occasional rebrightenings and spectral flattenings can be
associated with shocks, whether internal or external.

Scenario II -- the existence, sometimes, of a powerful jet from the
core in the soft state -- cannot be ruled out, but seems to be the
more complex solution, requiring explanations for why the jet spectrum
has switched to optically thin, and why it only happens sometimes.  

We conclude that the scenario I, which agrees with the current
paradigm of a suppressed (`quenched') core jet in the soft state is
more likely, although scenario II cannot be ruled out. In order to
test this further, sensitive high resolution (VLBI) and/or high
frequency (mm/IR) observations are required throughout an outburst.
Note that the near-IR evidence for suppression of the jet in the soft
state (Homan et al. 2005, Russell et al. 2007) clearly supports the
case that the flat-spectrum jet is suppressed, and therefore that the
jet power is significantly reduced (whether the jet emission is
suppressed at all frequencies or just becomes optically thin).

\section{When does the steady jet reactivate -- is the `jet line vertical' ?}

It is clear from many radio observations (e.g. Tanabaum et al. 1974;
Fender et al. 1999; Corbel et al. 2000, 2001; Gallo, Fender \& Pooley
2003) that the radio emission from BHXRBs reactivates when the sources
transition back to the hard state following a period in a softer
state. In FBG04 it was suggested (speculatively) that there may be a
vertical `jet line' for each HID (and maybe each source). In harder
states than this jet line a steady jet would be produced, and in
softer states the jet suppressed. Transition across the jet line from
harder to softer states (i.e. right to left) would result in a radio
flare (due to internal shocks as jet velocity increased), whereas the
transition from soft to hard (left to right; at lower luminosity)
would simply produce the reactivation of the (flat spectrum) jet
without a large flare.

It seems clear that the jet does remain on until at least the moment
of the radio flare , i.e. in the hard-intermediate state, during the
overall hard $\rightarrow$ soft transition (see Fig 1; also FBG04 and
Corbel et al. 2004). However, data testing the reactivation of the
hard state steady jet during the soft to hard transition has been hard
to come by.

\subsection{GX 339-4: the jet line does not seem to be vertical}

\begin{figure}
\centerline{\epsfig{file=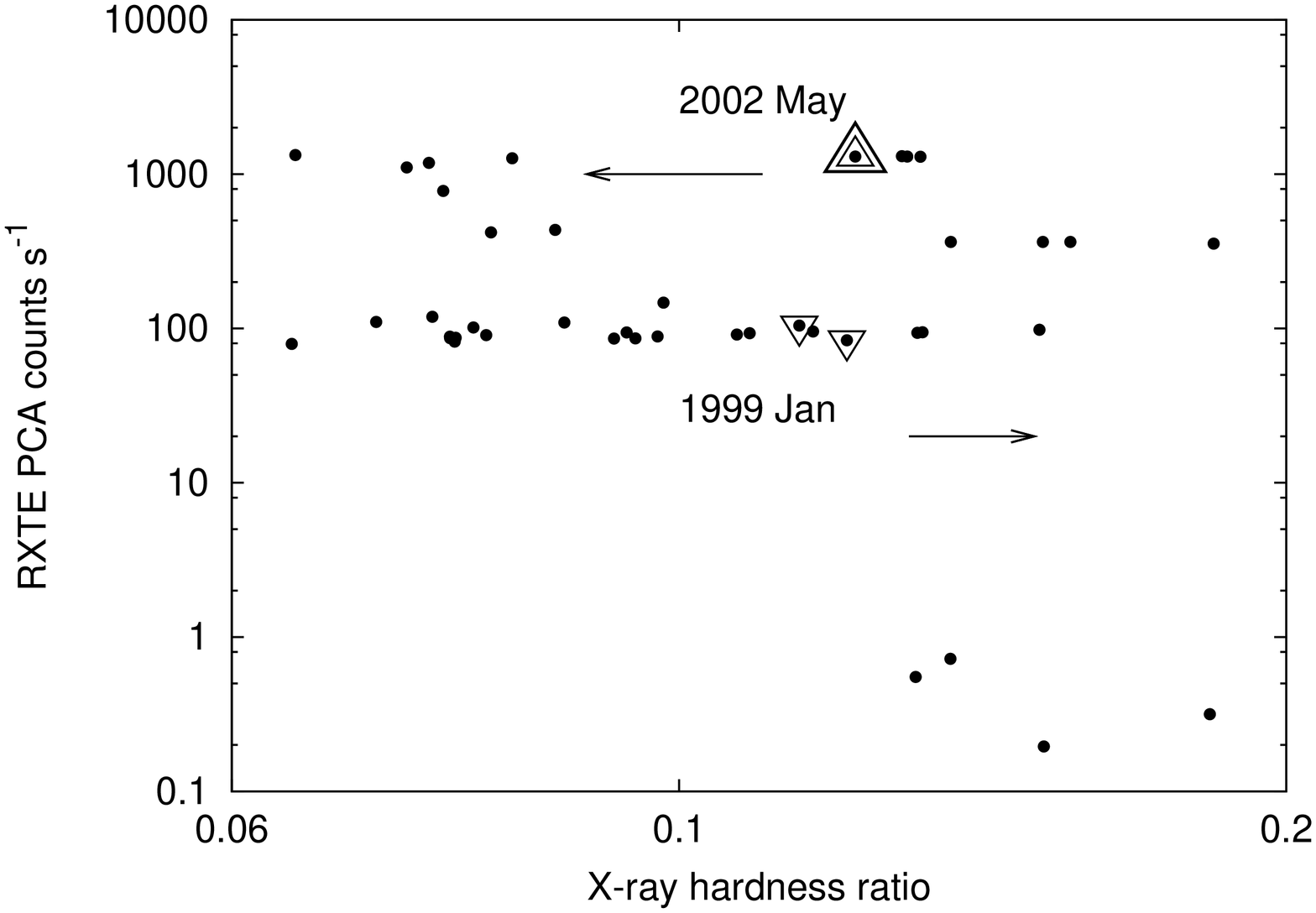, width=7cm, angle=270}}
\caption{Varying radio response at the same hardness in GX 339-4. In
  the 2002--03 outburst the source was detected at a steady level of
  $\sim 12$ mJy before flaring to $> 40$ mJy. Earlier observations, in
  January 1999, placed sub-mJy upper limits on the radio flux at the
  same hardness. The arrows indicate the general direction of motion
  in the HID at the time of the observations: in 1999 the source was
  returning to the hard state; in 2002 it was transiting from hard to
  soft.}
\label{339zoom}
\end{figure}

Comparison of the 1996--1999 and 2002--2003 outbursts of GX 339-4
indicates that upper limits to the jet flux have been measured, in
1999, at almost exactly the same hardness ratio where the major radio
flare was observed in 2002. This is shown in some detail in Fig
{\ref{339zoom}}. Note that there may be some uncertainties associated
with comparing colours at different epochs due to changes in the
response of the RXTE PCAs.

The upper limits to the radio flux density in 1999 are clearly also,
however, at much lower X-ray count rates. Could this be an explanation
for the non-detection ? The upper limits of $\leq 0.2$ mJy were
obtained at a count rate of $\sim 80$ ct/sec. The $\sim 12$ mJy steady
detection and subsequent flare were observed at a much higher count
rate, of $\sim 1300$ ct/sec. Combining these two results, a relation
between radio and X-ray fluxes steeper than $F_{\rm radio} \propto
F_{\rm X}^{1.4}$ would be required to account for the non-detection in
1999. This is of course far steeper than the $F_{\rm radio} \propto
F_{\rm X}^{\sim 0.7}$ relation of Corbel et al. (2002) for this and
other (Gallo et al. 2003) sources in the hard state. A +1.4 index is
in fact as expected for jets from a radiatively efficient accretion
flow (e.g. Heinz \& Sunyaev 2003; see Migliari \& Fender 2006 for the
possible case of such an index for neutron stars) and so this is not
quite ruled out. Nevertheless, it seems likely that this combination
of observations implies that a vertical `jet line', implying a
one-to-one correspondence between X-ray hardness and core jet
properties, does not exist. Of course there is no {\em a priori}
reason to assume a vertical jet line, but since this is what was
sketched in FBG04 we need to address this point.

As noted in Corbel et al. (2000), the first redetection of GX 339-4
after the recovery of the jet at the end of the 1999--2000 outburst
does have an unusually optically thin spectrum.

\subsection{XTE J1720-318: reactivation of the jet ?}

The single best example of jet reactivation in the hard intermediate
state appears to be the 2003 outburst of XTE J1720-318 (see Fig 1). In
this outburst the peak radio flux appears to occur at a hardness ratio
of $0.01 \leq H \leq 0.02$. During the soft to hard transition no
radio emission is detected to typical upper limits of 0.2--0.4 mJy to
a hardness of $\sim 0.02$, but radio emission is detected at a
hardness of $0.1 \leq H \leq 0.2$ 15 days later. This is clearly
softer than the hardest spectral state which seems to exist at $H \geq
0.3$ and in which the source remains in a further 30+ observations
over 120 days. In total there is a period of 69 days between radio
detections of the source. Two further detections are made as the
source settles back into the hardest spectral state.

\begin{figure}
\centerline{\epsfig{file=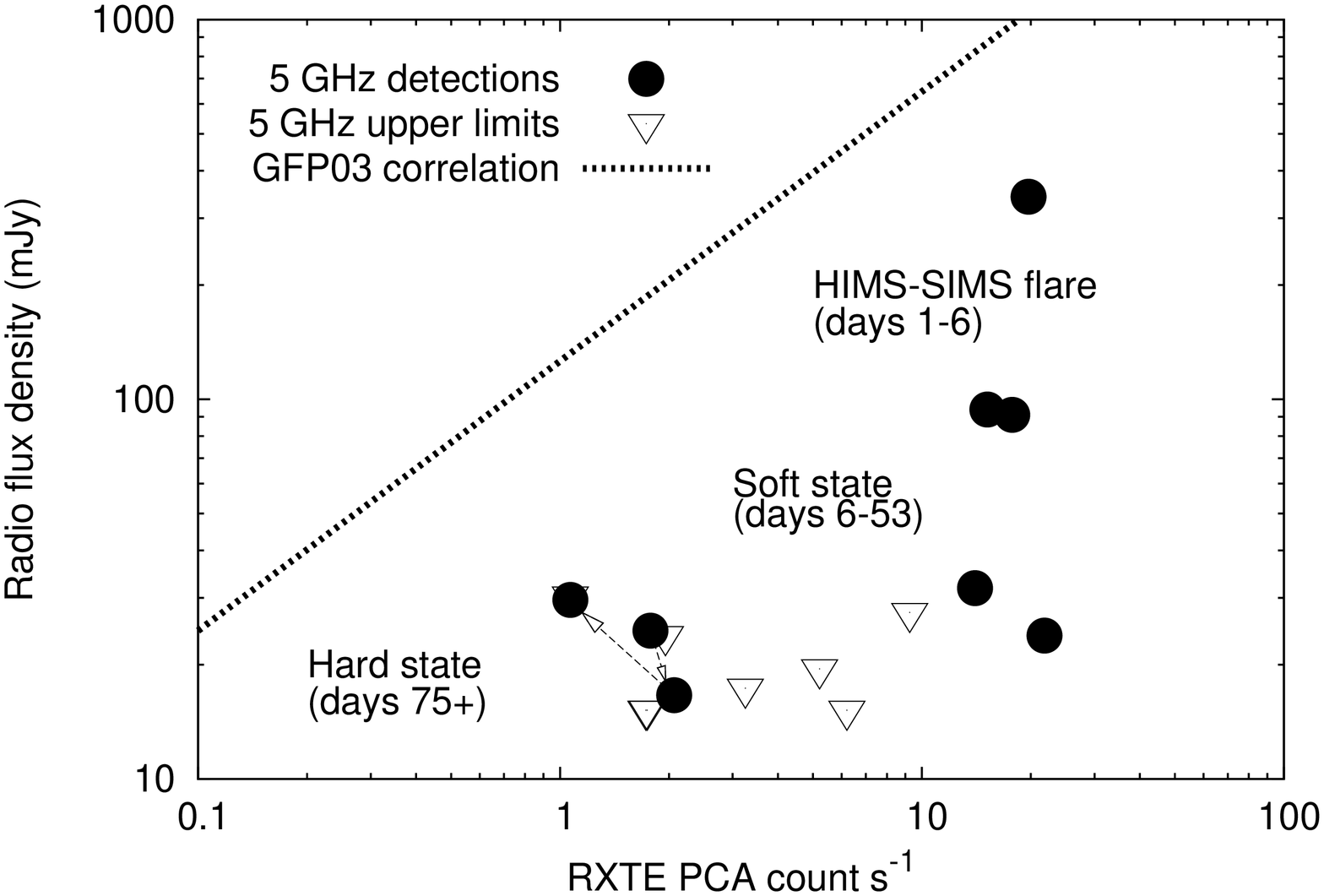, width=6cm, angle=270}}
\caption{Radio detections of XTE J1720-318 in the radio:X-ray plane,
compared to the overall hard state correlation from Gallo, Fender \&
Pooley 2003. The source is always below the relation (caveat distance
uncertainties) but appears to move back towards the relation during
reactivation on the lower hard-intermediate branch of the HID (see Fig
1), after a long period of non-detection in the soft state.}
\label{1720-r}
\end{figure}

Fig {\ref{1720-r}} plots the evolution of the XTE J1720-318 in the
radio:X-ray plane, with approximate indications of X-ray state and
also the correlation of GFP03 plotted.  If we compare the flux at this
radio reactivation with the $L_{\rm radio}$:$L_X$ relation presented
in Gallo, Fender \& Pooley (2003) and Gallo et al. (2006), we find
that the source is rather radio quiet both in outburst and in the hard
state (as noted already by Gallo 2007). However, the radio
reactivation does seem to bring the source closer to the GFP03
correlation.

%Compare to Belloni result for state transition, which might imply a
%diagonal jet line.

\section{Discussion}

In all cases investigated here -- 14 major outbursts compared to the
three studied in FBG04 -- the highest radio flux density occurred
during the hard to soft state transition near the start (and peak) of
the outburst. This seems to confirm the FBG04 assertion that major
ejections occur during the evolution from SEDs dominated by a
combination of corona plus jet, to those dominated by an optically
thick disc-like component. However, it has also become apparent that
in many systems the bright radio flare is associated with a phase of
X-ray flaring which is nowhere near as evident in our archetypal
system, GX 339-4. In some cases, e.g. XTE J1550-564, the disc can
appear to become extremely hot and physical changes are not well
mapped by the standard colours in the HID (as is the case for GRS
1915+105).

A key issue which arises in discussing the results compiled above is
how robust an indicator is the X-ray hardness of the physical state of
the accretion flow ? We may wish to compare the physical conditions of
the accretion flow at the moment of major ejection in different
systems, or compare the moments of jet flare and subsequent
suppression, with the moment of jet `reignition' on the lower
branch. In fact it is clear that on the lower (later) branch of the
outburst, at the same hardness (of course dependent on the energy
ranges used for the hardness ratio), the disc fraction can be very
different than on the upper (earlier) branch.  Dunn et al. (2008b)
present the relation between X-ray hardness and disc ($\equiv$ power
law) fraction (their Fig 3, right panel), illustrating the rather
complex relation between a given hardness ratio and the ratio of the
disc and power-law components.  Based on this discussion, it is clear
that the `jet line', {\em if it corresponds to a comparable ratio of
power-law to thermal components in the X-ray spectrum}, should not be
vertical in the HID (as sketched in FBG04) nor simply diagonal (as
sketched in Klein-Wolt \& van der Klis 2007) but rather more
complex. Disc-fraction luminosity diagrams (K\"ording, Jester \&
Fender 2006; Dunn et al. 2008 and in prep) or some other prescription
for describing the evolution of physical components, rather than
colours, may be required to make further progress here. 

\begin{table*}
\begin{small}
\begin{tabular}{lll}
\hline
Empirical aspect & Interpretation & Comments \\
\hline
Major radio flare during hard$\rightarrow$ transition & Major ejection & {\bf Confirmed}; additional X-ray flaring noted in some systems; \\
& & in one system five radio flares occur around same hardness \\
Reactivation of radio emission on return hard branch & Reactivation of jet (without flare) & {\bf Unclear}; only one useful example  \\
No radio emission in soft state & Jet 'off' in soft state & {\bf Consistent but not proven}; detections of (opt. thin) radio in \\ 
& & soft state; jet may be off ('remnant' radio) or intermittent \\ 
Low/hard state jet slower ($\Gamma \leq 2$) jet than & Leads to internal shocks & {\bf Unclear} Low/hard state jet velocity remains untested \\
subsequent powerful transient jet ($\Gamma \geq 2$) & & \\
Correspondence of power spectral changes with ejections & Ejection of corona ? & {\bf Unclear}; clear correspondence on timescales of days \\
& & but causal link not demonstrated \\
{\em Other aspects} & Jet power large & {\bf Confirmed}; many further studies appear to support this \\
& Varying inner disc radius at transitions & {\bf Unclear}; changes of order 10s of $R_G$ cannot be excluded \\
& Powerful ejection when disc reaches ISCO & {\bf Unclear}; some evidence against (e.g. XTE J1550-564) \\                      \hline
\end{tabular}
\end{small}
\caption{A brief tabular summary indicating aspects of the FBG04 model which have been tested, some empirical with interpretation, some mostly interpretation, some new, in this work.}
\label{summary}
\end{table*}

An interesting comparison case is Cyg X-1, where state transitions
occur in both directions at essentially the same luminosity ($\sim
2$\% Eddington). This suggests that if there really is any physical
connection between the disc:power-law ratio and the presence or
absence of a jet, it could be tested here.

It is somewhat disappointing at first to find that there appears to be
no one-to-one relation between abrupt changes in the rapid X-ray
variability properties, as measured in Fourier power spectra, and the
major radio ejection events. Both phenomena clearly happen around the
same time, but the case of GX339-4 (and possibly also XTE J1550-564),
in which the radio ejection appears to occur more than a day before
the r.m.s.-drop `zone' appears to rule out a direct causal
connection. We have briefly investigated the possible association
between ejections and different types of QPO (following the
classifcation of Casella et al. 2005) but the lack of a clear causal
link with zones of low-r.m.s. also rules out a one-to-one link between
QPOs and ejections. In particular it had been suggested (e.g. Soleri,
Belloni \& Casella 2008) that the type B QPO might be associated with
ejections, but the observation in GX 339-4 of a flare {\em prior} to
the appearance of these QPOs does not support this idea.  Before we
discuss further, some caveats are worth restating however:

\begin{itemize}
\item{As noted in FBG04, in the phase between the brightest canonical
hard state and the brightest ejection, there seems to be a phase of
instability in the jet, resulting in some flaring and rapid changes in
spectral index. When combined with the poor radio sampling much of our
conclusions are based upon, this remains a clear possibility for the
case of GX 339-4 -- i.e. there was an even larger flare which we
missed.}
\item{There are at least two forms of delay which will act to make the
radio emission appear later than the `trigger' X-ray event. Optical
depth effects (well understood since e.g. van der Laan 1966) will have
this effect, as will the delay required for two `shells' to collide in
the event that radio emission is generated by internal shocks
(e.g. Kaiser, Spruit \& Sunyaev 2000; Jamil, Fender \& Kaiser
2008). These delays are empirically observed in X-ray binaries to range
from minutes to days depending on the luminosity of the event and the
observing frequency.}
\end{itemize}

However, if these caveats are not relevant, then we are forced to
conclude that the major ejection events have an imprecise connection
to the X-ray `timing state' and possibly also to the spectral
state. In fact it is not clear how tight is the relation between the
two definitions of states (see discussions in e.g. Homan \& Belloni
2005; Remillard \& McClintock 2006; Klein-Wolt \& van der Klis
2007). This may imply that the jet, and maybe also the X-ray radiation
are common `symptoms' of some other change, but not themselves
causally connected. The bottom line is that current radio coverage is
simply not sufficient to pin down the moment of ejection to any
specific part of the complex X-ray transitions we have learned about
via RXTE.

\section{Conclusions}

In Fender, Belloni \& Gallo (2004) we presented a first attempt at a
unified picture for the radio:X-ray coupling in black hole X-ray
binaries. This picture was based upon the study of four black hole
systems, presenting one outburst each (one oscillation event in the
case of GRS 1915+105). One of the main conclusions of the paper was
that relativistic ejections are associated with the high luminosity
hard to soft X-ray state transition which occurs near the beginning of
most outbursts. In this paper we have investigated a far larger sample
of black hole X-ray binary outbursts observed by RXTE for which there
is at least some radio coverage. In all cases we find that the peak of
the radio emission, which is almost certainly associated with a
discrete relativistic ejection event (although there will be some
delay between ejection and radio peak), to be associated with the
overall hard to soft state transition, but we cannot pin down some
specific phase in the intermediate states. This confirms the result of
FBG04; however, we also note that in several systems the radio flare /
ejection event is associated with phases of X-ray flaring which are
not so evident in the most-studied source, GX 339-4.

We also find that in a large number of cases there is significant
radio emission in the soft state, where we have previously asserted
(Fender et al. 1999; FBG04) that the jet is suppressed or
`quenched'. It seems that in all cases this radio emission is
consistent with having an origin in jet-ISM interactions far from the
black hole, with the core radio emission indeed switched off. The
evidence for this comes in the form of optically thin radio spectra
and (usually) monotonic decays in radio flux in the soft state, as
well as the fact that in several sources the radio emission is indeed
strongly suppressed in the soft state. However, uncertainty about
exactly when the jet production mechanism shuts off remains, and is
unlikely to be resolved by flux monitoring observations with
relatively low angular resolution instruments such as ATCA and the VLA
-- higher resolution VLBI and/or higher frequency observations,
preferably simultaneous with X-ray observations, are going to be
necessary to make progress.

The reactivation of the core jet in the hard intermediate state was
predicted in FBG04 but there remains little direct evidence for this,
and it may well be that the core jet does not reactivate until the
canonical hard state is reached. The one tentative identification of a
hard intermediate state reactivation is the case of XTE J1720-318,
where a weak radio source appears during the transition back to the
hard state, but before the canonical hard state has been reached. More
radio observations of the decay phases of outburst are required to
investigate this further.

We have also attempted to extend our description of the X-ray
properties beyond just X-ray colours, but also to variability
properties. The bright intermediate states during which we have
clearly established the relativistic ejections take place are also
well known to be associated with the presence of strong QPOs, which
generally rise in frequency as the hard to soft transition progresses
(Klein-Wolt \& van der Klis 2007 and references therein). In
particular, we compared zones of anomalously low X-ray
r.m.s. variability, which are associated with transitions between
`hard intermediate' and `soft intermediate' states, with the times of
radio ejection events. These zones are {\em not} simply associated
with `disc dilution' of the variability signal, but by some as yet
unexplained reduction in the degree of variability of the hard X-ray
component; they are also often associated with type-B QPOs.  In all
cases where the data were good enough to measure both, the radio
flares and r.m.s. drops are coincident within a few days. In the case
of XTE J1859+226, there are hints that a sequence of five radio flares
are associated with a corresponding number of r.m.s. variability
drops, superposed on a general decline in the r.m.s. towards the soft
state. An obvious speculation would be that the radio flares are
associated with the ejection of the same coronal material which is
responsible for much of the variability (as suggested previously by
e.g. Rodriguez et al. 2003, Vadawale et al. 2003; FBG04; see also
Rodrigues \& Prat 2008). However, in the case of GX 339-4 it appears
that a strong radio flare event took place more than a day {\em
before} the X-ray r.m.s. drop, contrary to expectations for such a
scenario. Therefore at present it is uncertain whether the radio
flares and r.m.s. drops are simply independent symptoms of some other
underlying process, or that perhaps every dip is indeed causally
connected to a subsequent flare, but that our radio coverage is so
poor that we are missing most flares. Given the lack of a clear
one-to-one relation between the r.m.s. decreases and jet ejections, it
was clear that we would not (at this time) find a clear connection
between the presence of certain types of QPO and ejections.

We furthermore note that some aspects of the FBG04 model, both
empirical and interpretational, remain untested. Key amongst these is
the continued lack of a good measure of the speed of the jet in the
hard state. As noted in FBG04, and originally in GFP03 and Heinz \&
Merloni (2004), the relatively narrow range in normalisations of the
disc-jet coupling in the hard state of different black hole X-ray
binaries implies a narrow range of velocities for such jets. Although
we (still) favour a lower velocity, we are still lacking the single
good measurement of the velocity of a hard state jet in order to test
this. Measuring the velocity of a steady-state jet, relatively
low-power, jet will however require imaginative methods. The proposed
internal shock model suggested by a strong velocity differential, may
still be tested independently for black hole binaries as begun
recently by Jamil et al. 2008 (see also e.g. Pe\'er \& Casella 2009).
Several of the interpretational aspects of the FBG04 model also remain
untested, or even controverisal. Primary amongst these is whether or
not the inner disc radius really decreases in the hard $\rightarrow$
soft state transition. Several works (e.g. Miller et al. 2006; Rykoff
et al. 2007) argue against this interpretation, while others support
it sometimes using the same data (e.g. Gierlinski, Done \& Page 2008;
Cabanac et al. 2009). Table \ref{summary} summarizes which components
of the model have been tested by this study, and what aspects have
been added.

A key deficit in our understanding of the 'disc-jet' coupling in these
systems arises not from the lack of observations of the accretion flow
(i.e. X-rays), but from observations of the jet (radio, infrared),
particularly in comparison to studies of AGN, where there are a vast
number of radio observations. The next generation of radio facilities
(e.g. Fender 2008) will greatly improve on our coverage in this area,
with daily (maybe more frequent) radio observations of all active
X-ray sources. It is to be hoped that there will be an equivalent or
descendent of RXTE around in that time to provide the necessary
X-ray coverage.

\section*{Acknowledgements}

RPF would like to acknowledge assistance from James Miller-Jones, and
useful discussions with P Casella and M Klein-Wolt.  TMB acknowledges
support by grant IASI-INAF I/088/06/.  TMB and RPF acknowledge support
from the International Space Science Institute (ISSI), Team Number
116.

\end{document}